\documentclass[iop,apj]{emulateapj}

\usepackage{graphicx}
\usepackage{natbib}

\def\specchar#1{{\sc #1}}

\def\arcsec{\hbox{$^{\prime\prime}$}}

\def\CaII{\mbox{Ca\,\specchar{ii}}}

\def\HeI{\mbox{He\,\specchar{i}}}

\def\Halpha{\mbox{H$\alpha$}}

\def\kms{\hbox{km$\;$s$^{-1}$}}
\def\ms{\hbox{m$\;$s$^{-1}$}}

\def\pun{\stackrel{}{\mbox{.}}}
\def\farcs{$\stackrel{\prime\prime}{\pun}$}

\begin{document}


\title{Observational detection of drift velocity between ionized and neutral species in solar prominences}

\author{Elena Khomenko, Manuel Collados}
\affil{Instituto de Astrof\'isica de Canarias, 38205 La Laguna, Tenerife, Spain and \\
Departamento de Astrof\'{\i}sica, Universidad de La Laguna, 38205, La Laguna, Tenerife, Spain}
  
  \and
  
\author{Antonio J. D\'\i az }
\affil{Universitat de les Illes Balears, 07122, Crta Valldemossa, km 7.5, Palma de Mallorca, Spain}
\email{khomenko@iac.es, mcv@iac.es, aj.diaz@uib.es}

\shortauthors{Khomenko, Collados and D\'\i  az}

\shorttitle{Observational Detection of Drift Velocity}

\begin{abstract}\noindent

We report a detection of differences in ion and neutral velocities in prominences using high resolution spectral data obtained in September 2012 at the German Vacuum Tower Telescope (Observatorio del Teide, Tenerife). A time series of scans of a small portion of a solar prominence was obtained simultaneously with a high cadence using the lines of two elements with different ionization states, namely the \CaII\ 8542 \AA\ and the \HeI\ 10830 \AA. Displacements, widths and amplitudes of both lines were carefully compared to extract dynamical information about the plasma.  Many dynamical features are detected, such as counterstreaming flows, jets and propagating waves. In all the cases we find very strong correlation between the parameters extracted from the lines of both elements, confirming that both trace the same plasma. Nevertheless, we also find short-lived transients where this correlation is lost. These transients are associated with the ion-neutral drift velocities of the order of several hundred \ms. The patches of non-zero drift velocity show coherence on time-distance diagrams.

\end{abstract}

\keywords{Line: profiles, Sun: filaments, prominences}

\section{Introduction}

It has been stated for a long time that the solar atmospheric plasma is not in a neutral state nor in a fully ionized state. The effects of the partial ionization have been taken into account in many contexts, such as spectral line formation and inversion techniques. The influence of this plasma state on the dynamics becomes currently the focus of many investigations as well \citep{Arber2007, Soler2010,  Zaqarashvili+etal2012, Leake+etal2012, Khomenko+etal2014, Khomenko+etal2015, Cally+Khomenko2015}. Many studies of chromospheric and coronal plasma dynamics use the Magnetohydrodynamics (MHD) as main tool for successfully understanding the complex structure and dynamical processes of these solar atmospheric layers, but the standard MHD theory does not include partial ionization effects. An extended multi-fluid theory is a conceptually simple form to start integrating these effects without invoking mathematically more complex statistical approaches \citep[see][]{Braginskii, Bittencourt, Balescu}.

One of the basic assumptions of a multi-fluid theory is that the plasma is composed of different species, each of them behaving like a fluid and interacting with the rest of species, either directly or under certain conditions, with the collisional coupling between the fluids relatively weak. Hence, the plasma mostly behaves as a single fluid, but in certain processes there might be deviations between the dynamical and thermal properties of the different species, that are then smoothed by the interactions in relatively short time-scales (typically of the order of minutes).

The general transport equations for a multi-component plasma can be derived from the Boltzmann kinetic equation, taking into account general properties of the collisional terms \citep{Braginskii, Bittencourt, Balescu}. A basic assumption is that the collisional terms can be approximated as $\nu_{\alpha \beta} (\vec{v}_\alpha-\vec{v}_\beta)$, where $\nu_{\alpha \beta}$ is the collision frequency between the species labeled $\alpha$ and $\beta$ and $\vec{v}_x $ is the mean velocity of the species labelled $x$. Using this formalism, an equation for the relative velocity (referred as ``drift'' velocity in the rest of the paper) between ions and neutrals, namely $\vec{w}=\vec{v}_i-\vec{v}_n$, can be written as \citep{Diaz+etal2014, Khomenko+etal2014b}.

\begin{equation}
\label{eq:w}
\vec{w} = \frac{\xi_n}{\alpha_n} \left[\vec{J} \times\vec{B} \right] - \frac{\vec{G}}{\alpha_n} + m_e\nu_{en}\frac{\vec{J}}{e \alpha_n}
\end{equation}
In this equation $\xi_n$ is the neutral fraction; $\alpha_n$ is the sum of collisional frequencies between neutrals and other species multiplied by the corresponding mass densities; $\nu_{en}$ is the electron-neutral collisional frequency; $\vec{B}$ is the magnetic field vector, $\vec{J}$ is the current; $m_e$ and $e$ are the electron mass and charge. The quantity $\vec{G}=\xi_n \vec{\nabla}{p_{e}} - (1-\xi_n) \vec{\nabla}  {p_n}$ is the partial pressure gradient term, being $p_e$ the electron pressure and $p_n$ the neutral pressure. Equation \ref{eq:w} comes from combining the momentum equations for the electron, ion and neutral fluids, and neglecting the inertia terms compared to the friction terms. Its derivation is given in details in Section IV B of \citet{Khomenko+etal2014b}. Introducing this drift velocity in  the combined momentum equation of charged species, the induced electric field and the generalized induction equation can also be obtained \citep{Khomenko+etal2014b}.

\begin{figure*}
\centering
 \includegraphics[width=8.5cm]{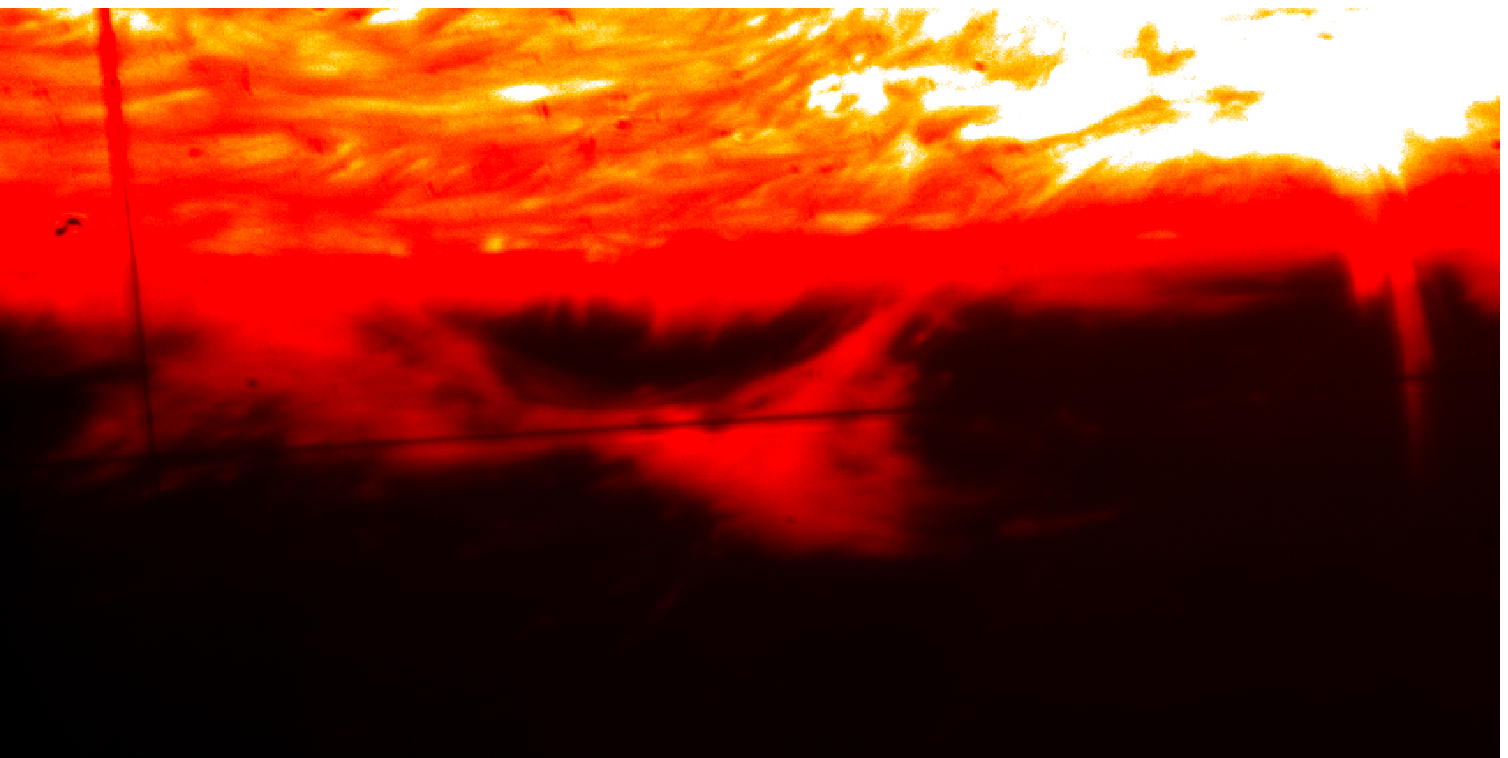}
 \includegraphics[width=8.5cm]{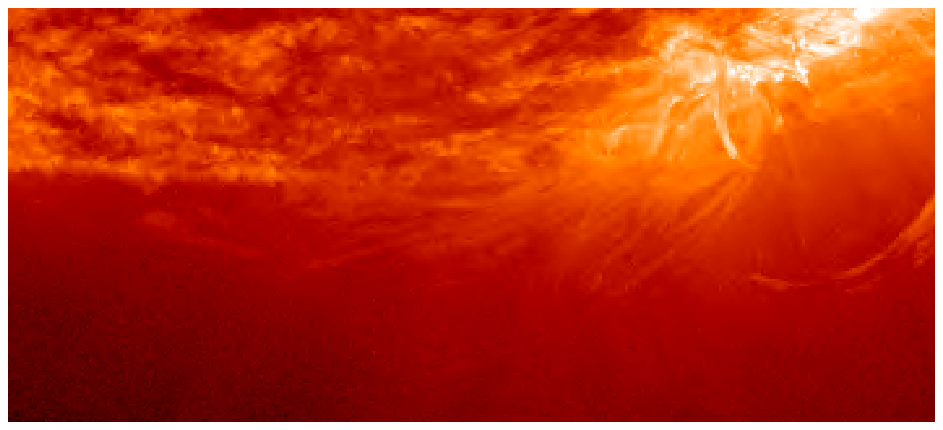}
\caption{Left: slitjaw image taken in the center of the \Halpha\ line. The horizontal black line marks the position of the slit, where the measurements in the \HeI\ and \CaII\ lines were taken. This particular image was chosen with the slit in the middle part of the 100th scan across the prominence, that is, 25 min after the start of the sequence. At this particular time, jets from the nearby active region (at the top right corner of the image) can be clearly seen). Right: image of the observed prominence taken by SDO 304 \AA\ channel. The extracted area is the same as in the left panel.}
\label{Halpha}
\end{figure*}

The existence of these drifts between species is thus a direct consequence of the partial ionization, and reflects that the coupling between the fluids is not strong enough to behave as a single fluid. However, in the physical conditions of the solar atmosphere, these terms are small, and electric fields and drift velocities are rapidly dissipated. The scale depends on the details of the process and the values of the physical parameters, but simulations show that the time scales involved are typically of the order of minutes or even less \citep{kc12, Khomenko+etal2014},  since the collisions are still efficient enough to prevent large deviations from single fluid theory.

To detect these effects, it is then necessary to measure as accurately as possible the velocity of different species at the same spatial position and simultaneously. Evaluating these drifts does not rely on unknown relations or assumptions, since measuring the Doppler shifts of different lines is straightforward. In a recent work, we made an attempt to detect differences between velocities of the Evershed flow measured co-spatially and simultaneously from spectral lines of neutral and ionized iron atoms \citep{Khomenko+etal2015}. The results reveal a slightly larger velocity of the Evershed flow in neutral lines. However, the drawback of such kind of measurements is the use of spectral lines with different formation heights in the photosphere of the Sun. On the one hand, there is an uncertainty regarding the precise formation height, depending on the model atmosphere taken and on the position on the limb, especially in an inhomogeneous environment such as a sunspot penumbra. On the one hand, the photosphere is dense and collisional effects dominate, which results in lower drift velocities expected according to Equation  \ref{eq:w}. In the present paper, we try to improve these two aspects. We choose as a target a solar prominence because its plasma can be considered relatively optically thin so that the formation region of different spectral lines largely occupies the same plasma volume. In addition, the physical conditions in prominences are expected to give rise to a significant partial ionization with a considerable amount of neutral and ionized species, depending upon the height in the prominence where measurements are made. Velocity fields are also non negligible and important mass flows \citep[e.g.,][]{Schmieder+etal1991, Zirker+etal1998, Lin+etal2003,  Lin+etal2005, Chae+etal2005, Chae2007, Chae+etal2008, Alexander+etal2013} and waves with different periodicities \citep[e.g.,][]{Oliver+Ballester2002, Banerjee+etal2007, Oliver2009, Tripathi+etal2009, Mackay+etal2010, Arregui+etal2012, Parenti2014}, as well as instabilities have been observed  \citep{Isobe+etal2005, Berger+etal2008, Heinzel+etal2008, Ryutova+etal2010, Berger2010, Berger+etal2011} . Instabilities may also lead to mass motions that can have a different impact on neutral and ions \citet{Soler+etal2012, Diaz+etal2012, Diaz+etal2014, Khomenko+etal2014}. Our main aim here is to investigate if any measurable drift velocity can be reliably detected in these structures and to discuss the implications of a possible detection in terms of partial ionization effects.

\begin{figure}[t]
  \centering
  \includegraphics[width=9cm]{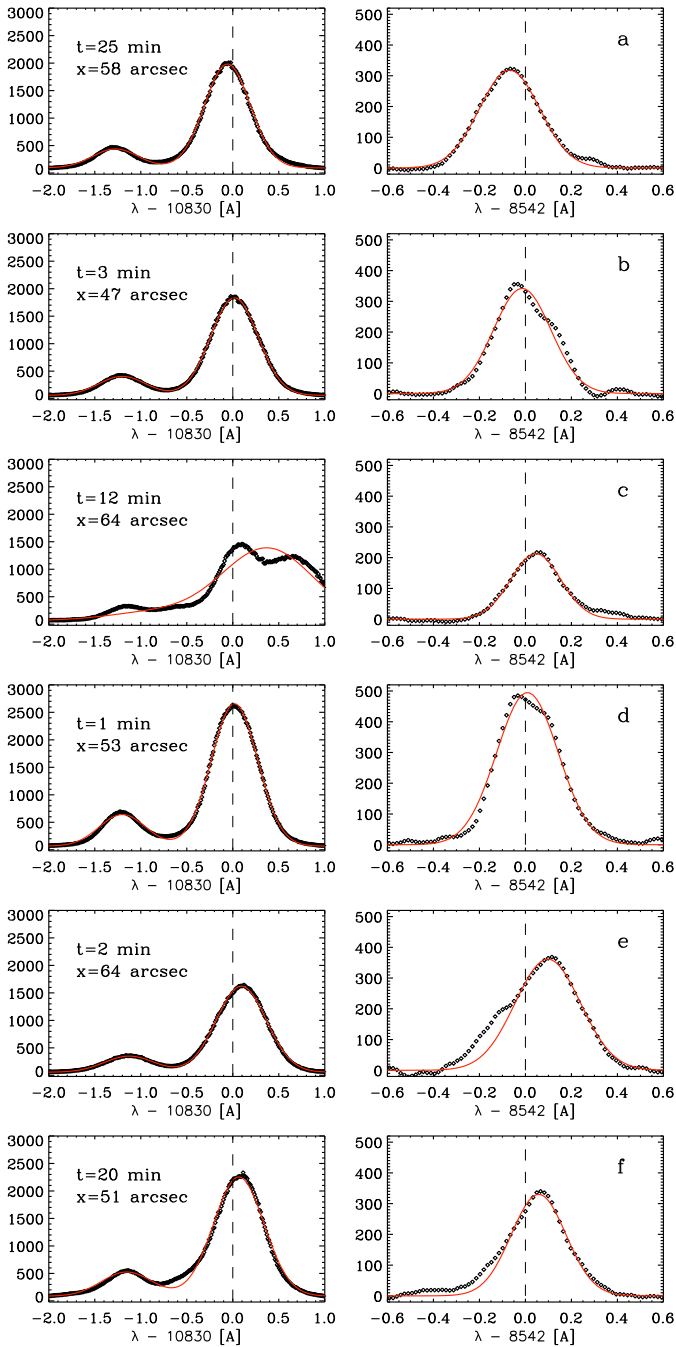} 
\caption{Pairs of spectra of the \HeI\ (left panels) and \CaII\ (right panels) lines for the six temporal and spatial locations along the slit shown in the panels. The red lines are the result of the single component fit. The profiles correspond to the middle of the scan (position 5) for both spectral lines. Horizontal dashed lines mark the positions of the average line centers.}
\label{fig:spectra}
\end{figure}

\section{Observational data}

\subsection{Observational details and target background}

The data used in this study were taken at the German Vacuum Tower Telescope (VTT, Observatorio del Teide, Tenerife) on  
2012 September 11. Several targets were registered during this campaign, but here we focus on one particular series where a prominence near AR11564 at S12W83 was observed continuously for more than half an hour with relatively good seeing conditions for the entire observation. The uncorrected $r_0$ parameter given by the adaptive optics (AO) software of the telescope \citep{Soltau+etal2002} gave values close to 10 cm during the whole series. The target was an active prominence undergoing evolution during the observation. It had well developed barbs, and at some parts of the slit there was some contamination from surges coming from the nearby active region that were superimposed with the prominence material. The AO system was working quite well locked in a nearby pore near the solar limb and did only jump once in the whole series.

We simultaneously detected spectra of \HeI\ 10830 \AA\ and \CaII\ 8542 \AA\  lines using the detector of the Tenerife Infrared Polarimeter \citep[TIP II][]{Collados2007} for the former and a camera optimized for the visible part of the spectrum for the latter. The spatial sampling along the slit was 0\farcs18, the same for both detectors. The time cadence was 1.5 sec per slit position for both spectral lines. To ensure such a high cadence and an acceptable noise level we did not use the polarimetric capabilities of TIP-II and only detected intensity spectra. The set has 1200 frames organized as follows: 120 scans of the body of the prominence in ten scanning positions separated by 0\farcs35 in the direction perpendicular to the slit. This means that an area of 3\farcs5 was observed and spectra at a fixed position were taken every 15 sec. Additionally, a slit-jaw \Halpha\ camera provided full field-of-view images of the prominence in the core of this spectral line. An example of an \Halpha\ image can be seen at the left panel of Figure \ref{Halpha}. Notice that these data were not used in this study anymore, since the camera did not provide spectral information about the \Halpha\ line. We only used \Halpha\ images to put the results from the other lines in context. 

Using the data from SDO 304 \AA\ channel (right panel of Figure \ref{Halpha}), we can see that the target prominence was close to the AR11564. The prominence itself can be seen in absorption on the disc during the previous days, and seems to be an active one related to the active region. The prominence spine was almost parallel to the limb and the prominence remained quite stable, despite the active region emitted jets during the observation. This particular active region did not produce any flare during the days before and after the observation.

The raw \CaII\ and \HeI\ spectra were reduced using a standard procedure. First of all, the data were cleaned from discrepancies between pixel counts from different parts of the CCD cameras by subtracting the averaged dark current images taken before and after the observational sequence. Then, the data were corrected for flat field  individually in each spectral region using the corresponding scans in \CaII\ and \HeI\  taken before and after the series at the solar disc center. The continuum of the flat field images was normalized to 10000 counts. The same normalization factor was applied to the prominence data, and the final amplitudes of the reduced spectra are in the same units of the disc center continuum for both lines. Finally, the only tip-tilt jump in the series was corrected by adequately displacing the spectral images in the slit direction. After these operations, data cubes of spectra of each of the two lines were obtained in every pixel along the slit and for any temporal position of the series. To perform the wavelength calibration, we compared the average flat field spectrum of the lines with the FTS atlas \citep{Neckel+Labs1984}. The spectral sampling of the \CaII\ spectra was 16.5 m\AA/pixel, and 11.0 m\AA/pixel for \HeI\ , and the spectral range was covered by 668 and 1010 pixels, respectively. 

The signal to noise level in the reduced spectra was about 11 in \CaII\ line and of about 400 in \HeI\ line.  The signal-to-noise level in \CaII\ line was relatively low because the emission is intrisically lower in this line and because of the low quantum efficiency of the camera used. The \CaII\ spectra were further filtered using a Principal Component Analysis technique  \citep{Rees+LopezAriste+Thatcher+Semel2000} which clearly improved the signal-to-noise ratio, as can be seen in Fig.\,\ref{fig:spectra}.  For that, each \CaII\ profile was represented by a linear combination of a set of eigenvectors $\vec{e_i}(\lambda)$ with appropriate constant coefficients $c_i$,
\begin{equation}
\vec{S}(\lambda)=\sum_{i=1}^n c_i \vec{e_i}(\lambda) \,.
\end{equation}
The system of eigenvectors was obtained from a dataset of randomly chosen 4000 \CaII\ profiles using a singular value decomposition (SVD) method  \citep{Rees+LopezAriste+Thatcher+Semel2000, Socas-Navarro+LopezAriste+Lites2001}.  In practice most of the eigenvectors do not carry information about the shape of the profiles, but only about the particular noise pattern of each profile. The truncation of the series allows therefore to remove the information about noise and to improve the signal no noise ratio. We truncated the expansion after the first 25 terms. This allowed to improve the signal to noise level  in \CaII\ by a factor of approximately 3.5.

\begin{figure*}[t]
  \centering
  \includegraphics[width=18cm]{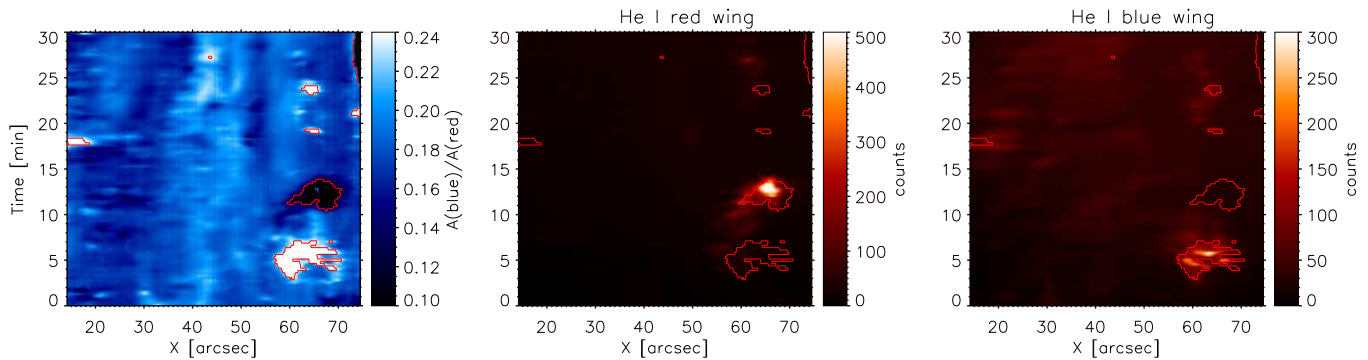} 
\caption{Ratio of the amplitudes of the blue to red components of \HeI\ triplet (left); intensity in the \HeI\ line at 1.4 \AA\ from the central position of the red component, corresponding to redshift velocity of 38.8 \kms (middle); intensity in the \HeI\ line at -1.96 \AA\ from the central position of the red component, corresponding to a blueshift velocity of 54.3 \kms (right). The contours mark the locations where the amplitude of the blue to red components of \HeI\ is either above 0.24 or below 0.12.  }
\label{fig:ratio}
\end{figure*}

\subsection{Data fitting procedure}

We fitted the observed \CaII\ and \HeI\ line profiles to obtain the physical parameters of the plasma necessary for our study. The structure of the \CaII\ spectra is generally relatively simple and consists of a single emission line profile. Therefore, we fitted the  \CaII\ line with a single Gaussian profile (with a base line that accounts for stray light coming from the disc), using the following equation

\begin{equation}
f(x)= a_0 \exp \left[ \frac{-(x-a_1)^2)}{2 a_2} \right] + a_3
\label{eq:Caprof}
\end{equation}

\noindent
with four degrees of freedom in the fit. Non-linear least square fits routines were used to find the best fit for these parameters, and the result is very good in over 80\% of the spectra. However, since the \CaII\ line counts were relatively low, at some pixels the noise led to problems in the fit. Hence, only those pixels  in which the maximum number of counts on at least one spectral position was over 5 times the noise level were selected and fitted. Because of this reason we excluded about 39\% of pixels where the prominence was not visible in  \CaII\  in the field of view. Such locations appear as black in the upper left and middle panels of  Figure \ref{fig:ajuste}. Multi-component profiles were also present at some time instants and locations, produced by surges and jets. The single-component fit failed for the profiles where multicomponent signals were clearly present. Such locations were identified and excluded from the analysis of velocities performed below. 

Regarding the helium triplet line, we chose as a typical line profile a two-Gaussian mold with fixed spectral separation and same width, and a with constant base line. The red components of the profile nearly overlap, so we have used a single Gaussian profile for them instead of resolving them separately \citep[as in, e.g.,][]{swd03, atl08}. We have not imposed fixed relative amplitudes between the elements of the triplet, therefore the ratio of the amplitudes of the blue and red components allows us to consider the opacity effects. Those effects are expected to be low for prominence material. The shape of the profile is then given by
\begin{eqnarray} \label{eq:Heprof}
f(x) &=& a_0 \exp{\left[\frac{-(x-a_1+\Delta)}{2 a_2}\right]} \nonumber \\
 &+& a_3 \exp{\left[\frac{-(x-a_1)}{2 a_2}\right]} + a_4,
\end{eqnarray}

\noindent
with five degrees of freedom. The spectral separation ($\Delta=1.219 \AA\ $) was obtained as the difference between the average positions of the red and blue components, see \citet{swd03}. As mentioned above, the number of counts is much higher in the \HeI\ line than in the \CaII\ line. Virtually, we could find a significant signal in all areas of the prominence. Nevertheless, the complicated structure of the triplet implied that in those cases where more than one contribution was present, the least-square fit could not distinguish if the signal had only one contribution of several of them (except in a few selected cases).

For both \CaII\ and \HeI\ lines the fits were weighted to give more relevance to spectral points with higher counts, since we are more interested in the peak position for velocity determination. 

In Figure~\ref{fig:spectra} we have plotted some spectra for the \HeI\ and \CaII\ lines taken at the same position and simultaneously. We have overplotted the result of the best fit in red lines. The first pair (panels $a$) corresponds to a regular point where the single component fit was very accurate; we can also check that in this particular example both spectral lines are blue-shifted. However, there are other spectra which show significant extra components in the red wing (panels $b$ and $d$) or the blue wing (panes $e$ and $f$) of \CaII.  In the cases as those shown in panels $b$ and $d$ the velocity shift between the components is relatively mild, while in the cases $e$ and $f$ is becomes more pronounced. At some rare instants there are components that only appear in one of the lines, such as happens at the panel $f$, where there is a discernible contribution in the \CaII\ blue wing, but no significant contribution in the \HeI\ triplet, although the secondary peak is slightly higher. This is related to the fact that \HeI\ line is much wider than the \CaII\ one and multiple components can be more easily distinguished in the latter case. There is also a prominent example of the contrary behavior, shown in the panel $c$ where the \HeI\ line shows a clear multiple component structure, while the \CaII\ signal is low amplitude and asymmetric with a discernible low-amplitude red component in its wing. Nevertheless,  in most of the cases both lines show the same features and thus we can conclude that the plasma emitting these lines is having very similar dynamical properties. Even if secondary components are present, their amplitudes in the majority of the cases are very close to the noise level.  The single component fit is accurate enough in over a 80\% of the points.

While single-component fit is acceptable in the cases shown in the panels $a$, $b$ and $d$ (with a somewhat larger width in the cases $b$ and $d$), the profiles as those shown in the panels $c$, $e$ and $f$ can not be reliably handled with such a fit. Their space time location is related to a plasma ejection from the nearby active region that happens to overlap with the prominence.

\subsection{Selection criteria}
\label{sect:selection}

Since in this work we are interested in reliably establishing the difference between the velocities of neutral and ionized atoms, we proceeded by discarding space-time locations where the velocities can not be measured reliably. Besides disregarding those points where the amplitude of the signal was too low, we used the additional following criteria to identify such locations: 
\begin{itemize}
\item Locations where the width obtained after the fit is above 6 \kms\  in \CaII\ line or above 12 \kms\ in \HeI\ line. The line width above these values reveals the presence of multiple components separated in velocity but not spectrally resolved. We excluded 2\% of pixels due to this criterion.

\item Locations where the width obtained after the fit is below 1 \kms\ in \CaII\ line, since an extremely low line width indicates errors of the fit. We also used this criterion to select locations where the signal in both \CaII\ and \HeI\ lines is reliably measured. This criterion coincides with the one based on the number of counts to identify the locations where the signal is above the noise in \CaII\ line. About 39\% of pixels were excluded.

\item Locations where the amplitude of the blue-wing or red-wing signal in \HeI\ is above a certain level. We choose as a reference the wavelengths 1.4 \AA\ and -1.96 \AA\ to the red and to the blue from the average line center position of the red component. The presence of a significant signal at these locations can be indicative of the existence of several components with large difference in velocities overlapping at the line of sight. 

\item Locations where the ratio of the amplitudes of the blue and red component of \HeI\ is above 0.24 or below 0.12. In the optically thin plasma this ratio should be exactly equal to 1/8=0.125. Larger values indicate that the plasma is not completely optically thin and lower values are not physically feasible. Since our aim is to measure \CaII\ and \HeI\ velocities originating from the same plasma, we discarded the locations where the plasma becomes thicker than a certain threshold, which would introduce an uncertainty about the location where the signal in both lines originates. About 5.5\% of points suffer this kind of problem in our data. Most of them are coincident with the criterion based on the amplitude of the blue-wing or red-wing signal in \HeI\ above.

\end{itemize}

Since in some of the pixels various items from listed above are present simultaneously, all in all we excluded about 44\% of pixels. The last criterion is the most important one in our analysis. Figure \ref{fig:ratio} shows the map of the ratio between the blue and red components of \HeI, $a_0/a_3$, see Eq. \ref{eq:Heprof}. It shows that the amplitude ratio varies around the mean value of about 0.18 for most of the prominence locations. The spatial variation of the ratio is in a narrow range staying essentially between 0.16 and 0.20. It indicates that the plasma in the observed prominence is very close to be optically thin with a very small spatial variation of the opacity. One distinguishes also some isolated locations where the ratio is above 0.24 or below 0.12. Those locations are marked by red contours and are excluded from the analysis. The middle and right panels of Fig. \ref{fig:ratio} shows the map of the \HeI\ signal in the red and blue wings, correspondingly. It can be seen that the locations with $a_0/a_3 > 0.24$ coincide with those of the strong blue-shifted signal (right panel) and those with $a_0/a_3 < 0.12$ coincide with the red-shifted signal (middle panel). This brings the last two criteria in the above list to be the same. 

\subsection{Errors of the fit and zero velocity reference}

In order to evaluate the velocity errors associated to the fit we proceed in the following way. The proper non-linear fitting routines gives a formal error of the fit of each of the free parameters in Eqs. \ref{eq:Caprof} and \ref{eq:Heprof}. However, those errors are very small and do not give a true evaluation of the uncertainty but a measure of the inaccuracy of the fitting procedure. We therefore assumed  that the uncertainties of the velocity measurements are limited by the wavelength resolution of observations. We took an upper estimate for such errors of half a pixel in wavelength. This gives us an uncertainty of $\pm 0.29$ \kms\ in \CaII\ and $\pm 0.15$ \kms\ in \HeI.

The wavelength calibration of the zero velocity position was done by comparing the averaged spectra of each of the lines to the FTS  atlas and correcting for the velocity of solar rotation at the latitude of the observed prominence, according to \citet{Snodgrass1984}. 

\begin{figure}[t]
  \centering
  \includegraphics[width=9cm]{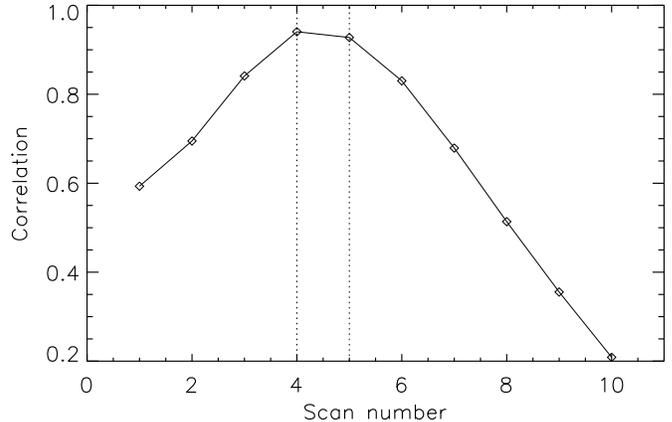}
\caption{Correlation coefficient between the time-slit map of the \CaII\ velocities from the 5th scan position, and the corresponding  time-slit maps of the \HeI\ velocities. The scan position of the \HeI\ velocities is given at the horizontal axis of the plot. }
\label{fig:correlation}
\end{figure}

\subsection{Differential refraction}

To be able to compare as accurately as possible the signal from the two lines, we need to take into account the differential refraction from the Earth atmosphere. The setup of the campaign ensured that the data from the two spectral regions were taken simultaneously and with the telescope pointing to the same place, but since both infrared lines have different wavelengths, they experiment a different refraction angle, so the light entering the optical system of the telescope does not originate from the same position in the Sun. Because of the orientation of the slit during the observation, this effect was almost in the direction of the scan.

A direct calculation (using the time of the day and the relative position of the Sun to the horizon) gives us a value of 0\farcs3. Additionally, we have compared the images from the different scanning positions in order to find the best correlation between them. In Figure~\ref{fig:correlation} we plot the correlation between the time-slit maps of velocities for different scanning positions. The best correlation is obtained when the \HeI\ map is taken at the previous scanning position (which is shifted 1.5 s in time), but the correlation is almost as good with the same position. This means that the differential refraction is present, but is below the size of the scanning step (0\farcs35). Since both positions give very similar results, for the rest of the paper we compare the results for the \CaII\ and \HeI\ lines at the same scanning positions because this implies no temporal shift. 

\begin{figure*}[t]
  \centering
  \includegraphics[width=18cm]{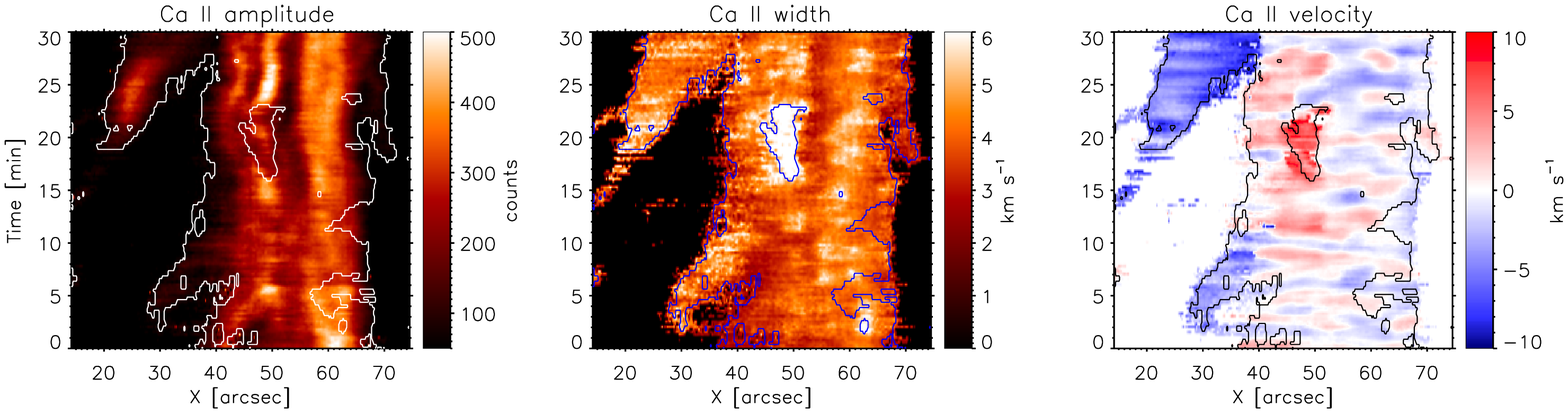}
  \includegraphics[width=18cm]{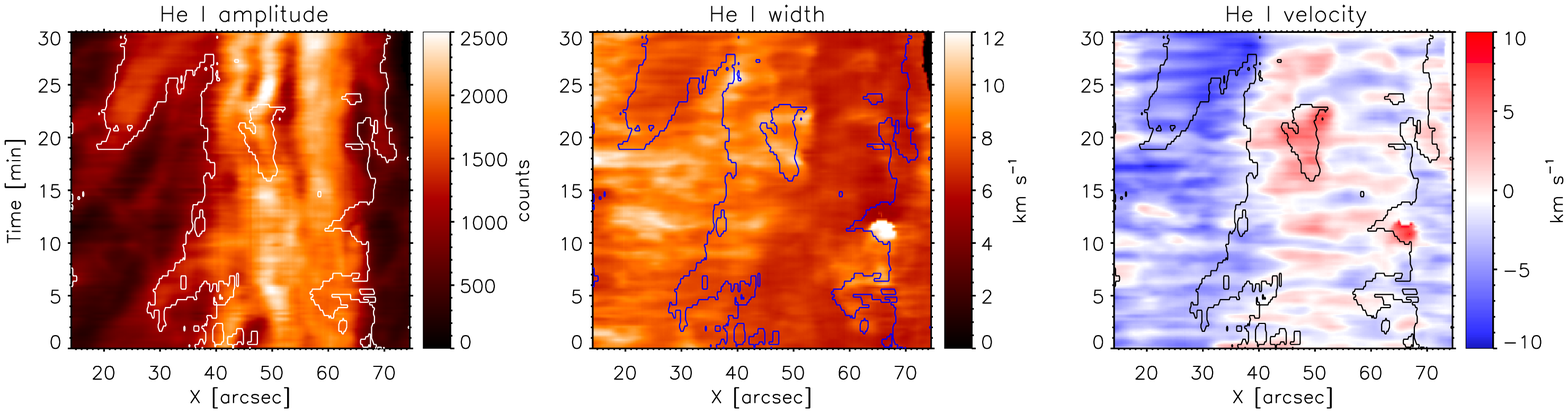}
\caption{Time-slit maps showing the parameters obtained after one-component Gaussian fit to the \CaII\ line (upper panels) and \HeI\ line (lower panels). Plots from left to right correspond to the amplitude $a_0$, the line width $a_2$ , and the Doppler displacement $a_1$. Contours underline the locations where the selection criteria are satisfied, see Sect. \ref{sect:selection}. }
\label{fig:ajuste}
\end{figure*}

\section{Results of the Gaussian fit}

\subsection{Plasma parameters}

We proceed to analyze the parameters resulting from the Gaussian fit. Figure~\ref{fig:ajuste} shows the time-slit maps of the amplitudes, widths and displacements for the 1st scan position for both \CaII\ (top) and \HeI\ (bottom) lines. Similar results are obtained for the rest of the scan positions. The contour lines (same in all panels) underline the locations selected under the criteria listed in Section \ref{sect:selection}. One can observe that the Gaussian amplitude, which can be used as a proxy of the density of the plasma, is quite uniform in the emitting regions, with some higher values in specific knots. The Doppler velocities show spatial and temporal correlations, with plasma flowing towards and away from us at a fixed slit position, depending on the time instant.

The comparison between the time-slit maps shows that both lines originate from plasma with very similar dynamical conditions. The amplitudes of the lines (left panels) are coherent and show the same features, taking into account that the signal in the \CaII\ line was much lower, meaning that we could not fit it confidently in those points with lower counts. Regarding the line widths (middle panels) we also verify that there is a good agreement between the lines, taking into account that the thermal width (assuming LTE emission) is related to the atomic number, so that the width of the \CaII\ line is related to the width of the \HeI\ line by a factor $(m_\mathrm{Ca}/m_\mathrm{He})^{1/2} \approx 3.16$. However, the most striking similarity appears in the Doppler velocity (right panels), where the same features can be identified in both lines.

\subsection{Dynamics of the prominence and nearby plasma}

There are several dynamical processes that were taking place during the observations of the prominence. The slit was crossing the body of the prominence, while at some moments the jets from the nearby active region were contaminating at the line of sight.

\subsubsection{Waves}

One can see in Figure~\ref{fig:ajuste} that there is strong evidence of wave behavior in the observed signal. Since the information about the magnetic field vector is not available, it is hard to infer the type of wave. Nevertheless, the wave parameters can be inferred rather reliably. 

The wave main period of the waves is around 3 min. The 3 min oscillations are present almost during all the observation at all the slit positions covering the prominence. The wavelength of oscillations appears to be larger that the whole observed prominence (more than 60\arcsec), i.e. the whole body of the prominence is oscillating, although not necessarily in phase at all the parts. The amplitude of oscillations reaches 3 \kms\ and the oscillations appear to be slightly non-linear. 
The oscillations continue  without apparent damping during all the observation for more than half an hour.

\subsubsection{Active Region Jets}

The active region close to the observed prominence showed some activity, with jets and plumes sometimes coming from it and crossing the slit. Most of this activity lied in the other part of the slit not shown in Fig.  \ref{fig:ajuste}. However, at one instance we found evidence of one of these jets in the spectral information where prominence signal was also recorded. An example of such spectra is the pair $c$ from Fig.~\ref{fig:spectra}. The space-time evolution of the jet can be followed in Fig. \ref{fig:ajuste} at time around 12 min and location between 60\arcsec\ and 70\arcsec. It can also be seen in the \Halpha\ slitjaw images.The line-of-sight velocity component is almost zero at the onset of the jet, but it becomes much higher at later stages, reaching over 50 \kms. The jet is clearly seen in the red wing, and it accelerates as time passes until its amplitude vanishes without slowing down or reversing the velocity sign.

\begin{figure*}[t]
  \centering
  \includegraphics[width=18cm]{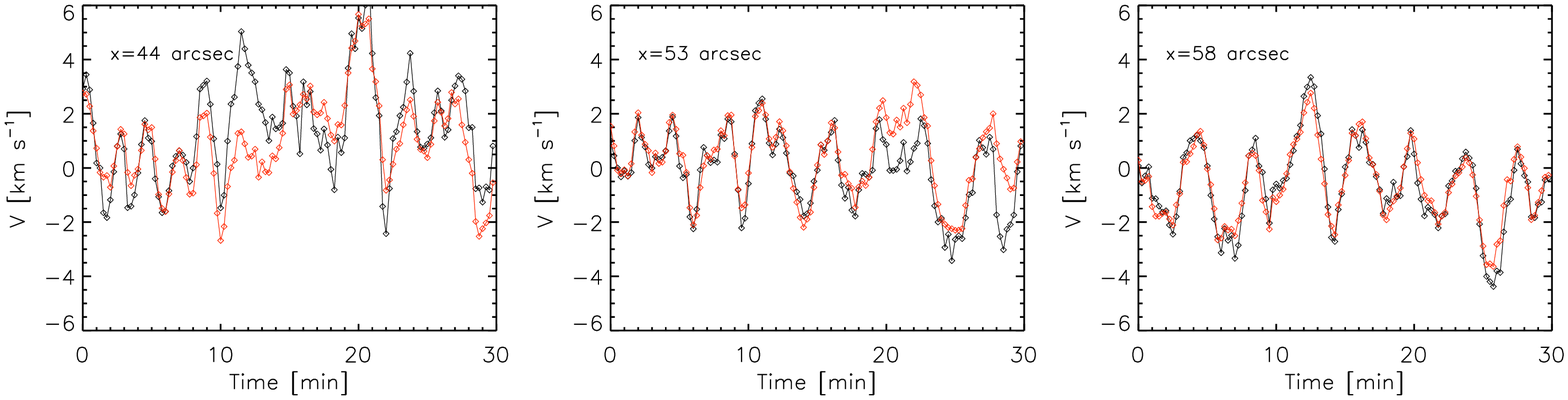}
  \includegraphics[width=18cm]{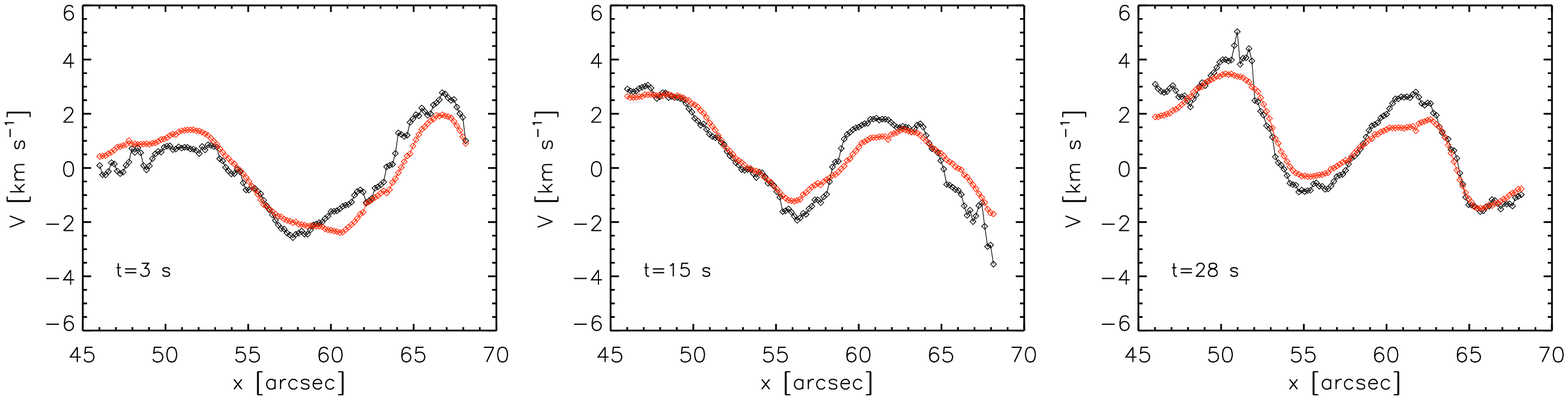}
\caption{Upper panels: time series of velocity variations for a fixed position along the slit. Bottom panels: spatial variations of velocity along the slit at a fixed time. The results for the 1st scan position are shown.  Black lines are \CaII\ velocities, red lines are \HeI\ velocities. }
\label{fig:cuts}
\end{figure*}

\begin{figure*}[t]
  \centering
  \includegraphics[width=18cm]{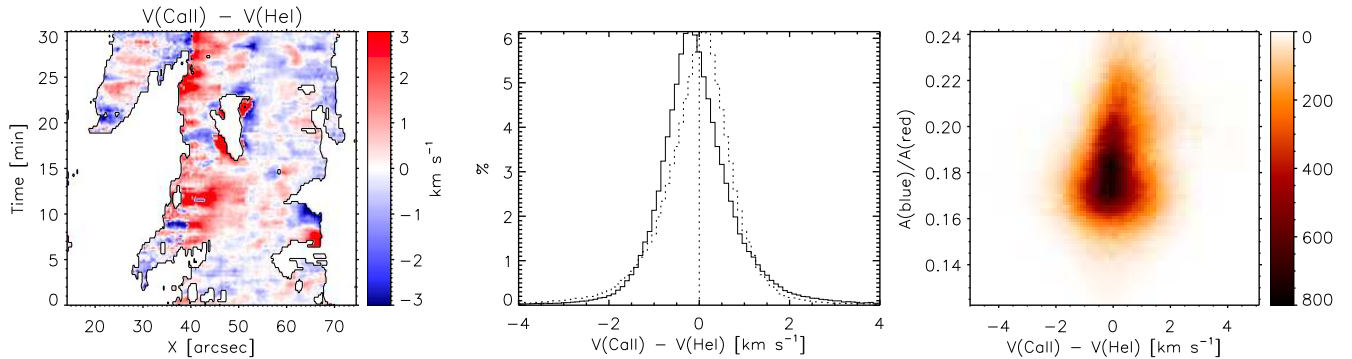}
\caption{Left: Drift velocity, $w=v_{\rm CaII} - v_{\rm HeI}$ , as a function of time and distance along the slit. Contours underline the locations where the selection criteria are satisfied. Outside these locations the velocity difference is set to zero. The results for the for the 1st scan are shown. Middle: histogram of $w$  (solid line) and the histogram of $-w$ (dotted line) at the selected locations for all scans. Right: bi-dimensional histogram showing the dependence between the ratio of the amplitudes of blue to red \HeI\ components and  \CaII\ - \HeI\ velocity difference at the selected locations for all scans.  }
\label{fig:hist}
\end{figure*}

\begin{figure*}[t]
  \centering
  \includegraphics[width=18cm]{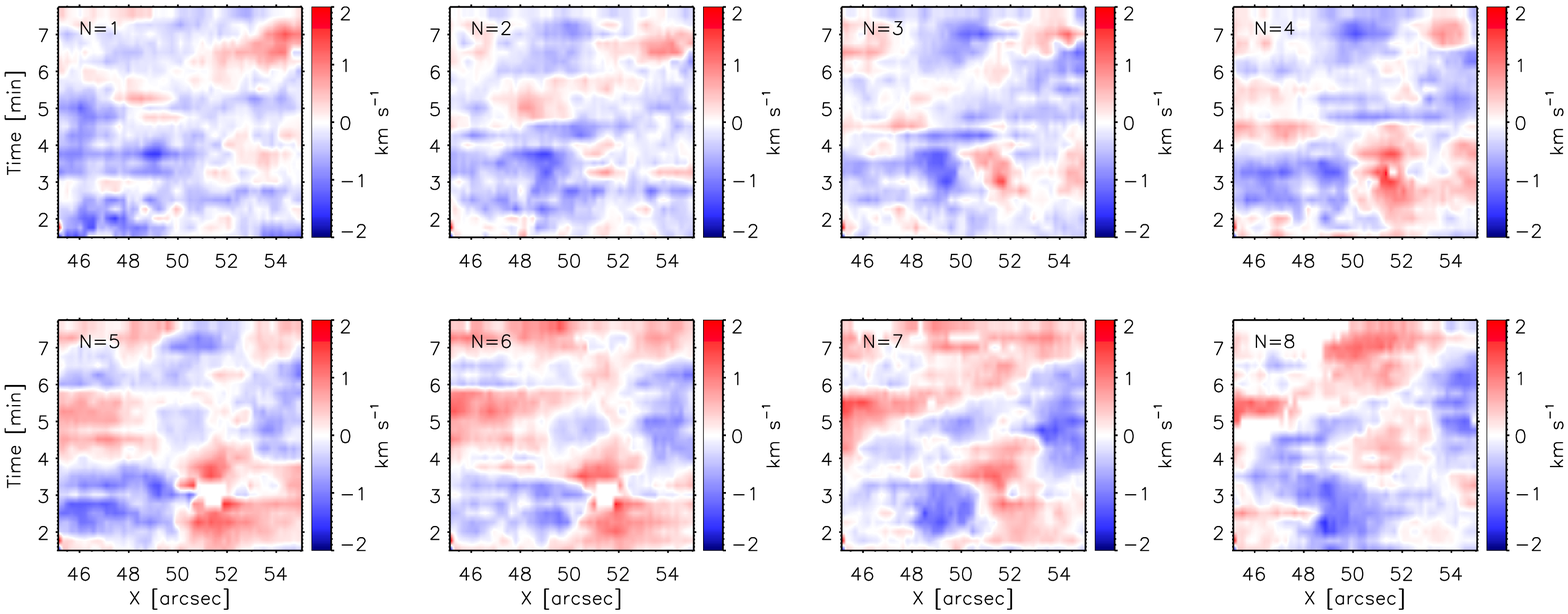}
\caption{Zoom of the time-slit map of the difference between \CaII\ and \HeI\ velocity showing a small area where the one-component fit is reliable for 8 scanning positions of the slit, indicated in the figure. }
\label{fig:slits}
\end{figure*}

\section{Detection of drift velocity}
\label{sect:drift}

In this section we compare in detail the velocities obtained from the \HeI\ and \CaII\ lines shown in Fig. \ref{fig:ajuste}. We assumed that the \CaII\ line serves as a proxy of the movement of ionized species, while the \HeI\ line is related to neutral species. The drift velocity is defined as the difference between the velocity of both lines, $w=v_{\rm CaII} - v_{\rm HeI}$ . 

Figure~\ref{fig:cuts} provides examples of  time and space cuts through the velocity maps from Fig. \ref{fig:ajuste} for both spectral lines. The upper panels of this Figure shows that the match between the velocities of the lines is extremely good. Both lines follow closely each other and show the same displacements, with very little differences between them. Nevertheless, at the instants of extreme velocities (maxima and minima) there is some difference in the behavior of both lines, with \CaII\ velocities being slightly larger. We could not find evidences that the difference between \CaII\ and \HeI\ velocity follows the 3 min oscillations, as individual velocities do. The phase shift between the velocities of both lines is found to be around zero. 

Similar behavior is also found when considering spatial variations at a fixed time (bottom panels of Fig. \ref{fig:cuts}). There, one may observe similar patterns for both lines, with some hints that spatial variations of \HeI\ velocities are slightly smoother. This is particularly evident at the bottom right panel of Fig. \ref{fig:cuts} where the gradients of \CaII\ velocity are slightly more pronounced. 

The left panel of Figure~\ref{fig:hist} shows the time-slit map of the drift velocity $w=v_{\rm CaII} - v_{\rm HeI}$  for the same scan position as in Fig. \ref{fig:ajuste}, obtained by directly subtracting the velocity maps from that figure. We set $w$ artificially to zero at locations that do not satisfy the selection criteria outlined in Sect. \ref{sect:selection}.

The inspection of this figure confirms the conclusion already apparent from Fig. \ref{fig:ajuste} that the difference between \CaII\ and \HeI\ velocities is small in most of the locations, taking into account the error bars. However, patches of blue and red colors are distinct, showing regions where the drift velocities are non-zero. Typical values at those locations are in the range of $\pm 1$ \kms. There are also patches where $w$ is positive and above 2 \kms\ located at the left border of the structure visible in the image. The inspection of the line profiles of both lines at these locations reveal that they have some asymmetry. The latter may be a consequence of the line of sight velocity gradient  at these locations, affecting the velocity measurements by the method adopted in this paper. 

The middle panel of Fig. \ref{fig:hist} shows the histogram of $w$ over all selected locations of the time-slit map and all scans (solid line). The dotted line in the same figure is the histogram of the same quantity but with the opposite sign, shown for the purposes of highlighting the asymmetry of the distribution.  It can be seen that the distribution of the relative velocity is slightly asymmetric, with its most probable value being slightly negative (i.e. \HeI\ velocity larger than \CaII\ velocity), but with a more extended tail toward the larger positive values (i.e. more locations where \CaII\ velocity is significantly larger than \HeI\ velocity). This histogram confirms the impression from Fig. \ref{fig:cuts} where we have observed that \CaII\ shows larger extreme values of the velocity compared to \HeI. The average value of $w$ over all locations is very close to zero.

We have verified whether the values of $w$ are affected by the opacity of the prominence material, i.e. if the difference in \CaII\ and \HeI\ velocity origins because the lines are not formed as exactly the same location. The right panel of Fig. \ref{fig:hist} shows the bi-dimensional histogram of $w$ as a function of the ratio between the blue to red amplitudes of \HeI\ profile, as an indicator of the opacity of the prominence. It reveals no dependence between both quantities. Therefore, we conclude that, to the first order, the presence of non-zero $w$ is not due to line formation effects. 

We also checked whether there is a dependence between $w$ and any other line parameters, such as the Doppler width, amplitude or displacement. The bi-dimensional histograms of the kind as the one shown in Fig. \ref{fig:hist} reveal no such dependence for any of the quantities. Nevertheless, the locations with non zero in $v_{\rm CaII} - v_{\rm HeI}$  are not randomly distributed over the time and space, but there is  temporal and spatial coherence. The areas with non-zero $w$ cover about 2\arcsec\ in space and have a typical lifetimes around 1 minute.

Some of those areas may correspond to locations where jets  contaminated the signal from the prominence at the field of view. Despite we have tried to avoid the areas with multi-component profiles, it is still possible that the selection criteria used do not completely eliminate such locations.  Nevertheless, at other locations the one-component fit is reliable but still $w$ is non zero. This is the case in most of the time-slit map. 

We analyzed with more detail one of the regions in Figure~\ref{fig:slits} with reliable one-component fit. The spectra in this area have no significant second components, so the fits are quite reliable and the effect can not be attributed to problems in the fit. The plots show that there is part of the prominence with patches of non-zero drift velocity distributed coherently both across the slit (with the typical the size of about 2\arcsec$-$3\arcsec) and in time (with lifetimes about $1-2$ minutes). The structures coherently evolve from one slit location to another. The fact that coherence is maintained only for short periods of time below one minute reinforce the necessity of the high time- and spatial resolution observations in order to reliably detect those drift velocities.

\section{Discusion and conclusions}

In this paper we have analyzed high temporal and spatial resolution observations of a prominence done simultaneously in an ionized \CaII\ line and a neutral \HeI\ line. Our analysis reveals that the structures observed with both lines are very similar indicating that they both form in essentially the same plasma volume. The velocities obtained from both lines are also very similar. At the same time, we obtained that there are small scale and small life time transients where the balance between both velocities is broken and drift velocity is observed. The balance is usually lost at places with large individual velocities or large spatial or temporal gradients. The magnitude of this drift velocity is below 1 \kms\ in most of the locations where it is detected, and both positive and negative values of $w$ are measured, apparently unrelated to the dynamical processes (such as wave motions) that the prominence was undergoing during the time of observations. The patches with non-zero drift velocity are distributed coherently in time and space when carefully considering only cases with reliable one-component fit to the profiles. Large drift velocities are also detected at locations where some jet from the nearby active region overlaps the observed field of view. However, at those locations, multi-component profiles are typical for one or both spectral lines, and the one-component fit becomes less reliable. Those locations are nevertheless interesting and need further detailed investigation.

There are several effects than may be responsible for the appearance of the mismatch between the velocities of both spectral lines. As mentioned in the introduction, in a partially ionized atmosphere, as collisions weaken, the ionized and the neutral plasma components become partially decoupled. The drift velocity given by Eq. \ref{eq:w} depends on currents, magnetic field and partial pressure gradients of the species. Our finding from Fig. \ref{fig:cuts} that the mismatch between \CaII\ and \HeI\ velocities becomes more pronounced at locations with larger individual velocities and larger gradients may serve as a confirmation that the detected drift velocities are due to physical decoupling of the components by partial ionization effects. The fact that the patches are coherent over space and time, are short-lived and only occupy small areas also provides confirmation that the observed effect are due to some physical process rather than observational drawbacks. Numerical simulations of prominence instabilities by \citet{Khomenko+etal2014} show that similar amplitudes of the drift velocities are expected in the prominence-corona transition region.

Other explanations for the mismatch between \CaII\ and \HeI\ velocities are also possible. One of the possible drawbacks of our approach is the uncertainty about the formation region of both spectral lines. The prominence material is frequently assumed to be optically thin. The ratio between the amplitudes of the blue and red components of \HeI\ profile shows that the observed prominence plasma was indeed very close to optically thin, with some slight variations of the opacity in space in time. However, we have found no correlation between the amplitude ratio of the \HeI\ components and the magnitude of the drift, see Fig. \ref{fig:hist}. Figure \ref{fig:ajuste} demonstrates that the amplitudes and widths of both lines are well correlated, taking into account the fewer counts in the \CaII\ line. Therefore, it can be concluded that the velocity signal measured by both lines originates at essentially the same locations over most of the observed prominence.  

We were able to detect drift velocities due to the very high temporal resolution of our observations. If the signal would be integrated over larger intervals, the effect apparent in Figures \ref{fig:cuts}, \ref{fig:hist} and \ref{fig:slits} would probably be lost or become much smaller. It is then expected that, if the resolution increases, the amplitudes of the ion-neutral velocity difference would become larger. In future works, it would be desirable to include more ionized and neutral spectral lines for the analysis in order to confirm the physical origin of the non-zero drift velocities measured in this work.

\begin{acknowledgements}
This work is partially supported by the Spanish Ministry of Science through projects AYA2010-18029, AYA2011-24808 and AYA2014-55078-P. This work contributes to the deliverables identified in FP7 European Research Council grant agreement 277829, ``Magnetic connectivity through the Solar Partially Ionized Atmosphere''. The authors also thank Drs. M. Luna, R. Oliver and M. Zapior for discussions.
\end{acknowledgements}

\end{document}